\documentclass[prb,twocolumn,nofootinbib,superscriptaddress,amssymb]{revtex4-1}
\usepackage{natbib}
\usepackage{graphicx}
\usepackage{dcolumn}
\usepackage{amsmath,amssymb,amsthm}
\usepackage{color}

\begin{document}

\title{ Back-action effects in cavity-coupled quantum conductors}

\author{Valeriu Moldoveanu}
\affiliation{National Institute of Materials Physics, Atomistilor 405A, Magurele 077125, Romania}
\affiliation{Centre International de Formation et de Recherche Avanc\'{e}es en Physique, Atomistilor 407, Magurele 077125, Romania}
\author{Ion Viorel Dinu} 
\affiliation{National Institute of Materials Physics, Atomistilor 405A, Magurele 077125, Romania}
\affiliation{Centre International de Formation et de Recherche Avanc\'{e}es en Physique, Atomistilor 407, Magurele 077125, Romania}
\author{Andrei Manolescu}
\affiliation{School of Science and Engineering, Reykjavik University, Menntavegur 1, IS-101 Reykjavik, Iceland}
\author{Vidar Gudmundsson}
\affiliation{Science Institute, University of Iceland, Dunhaga 3, IS-107 Reykjavik, Iceland}
\date{\today}

\begin{abstract}

We study the electronic transport through a pair of distant nanosystems ($S_a$ and $S_b$) embedded in a single-mode cavity. 
Each system is connected to source and drain particle reservoirs and the electron-photon coupling is described by the  
Tavis-Cummings model. The generalized master equation approach provides the reduced density operator of the double-system 
in the dressed-states basis. It is shown that the photon-mediated coupling
between the two subsystems leaves a signature on their transient and steady-state currents. In particular, a suitable bias 
applied on subsystem $S_b$ induces a photon-assisted current in the other subsystem $S_a$ which is otherwise in the 
Coulomb blockade. We also predict that a transient current passing through one subsystem triggers a charge transfer 
between the optically active levels of the second subsystem even if the latter is not connected to the leads. 
As a result of back-action, the transient current through the open system develops Rabi oscillations (ROs) 
whose period depends on the initial state of the closed system. 

\end{abstract}


\maketitle

\section{Introduction}
Promising applications of cavity quantum electrodynamics (QED) to spintronics are essentially rooted in the 
entangled dynamics of hybrid light-matter nanosystems. The presence of long-range electromagnetic coupling 
between distant quantum systems (ideally viewed as two qubits) has already been confirmed in self-assembled 
double quantum dots \cite{Laucht,Gallardo} and color centers embedded in photonic-crystal cavities \cite{Evans662}. 
In the field of circuit-QED Fink {\it et al.} \cite{Fink} reported that the optical transmission spectrum of a 
superconducting qubits array embedded in a microwave resonator can be explained by relying on a Tavis-Cummings-Dicke 
 Hamiltonian \cite{TC,PhysRev.93.99,doi:10.1098/rsta.2010.0333}. 

The Tavis-Cummings (TC) model provides the luminescence spectra and lasing properties of $N$ 
two-level systems (TLS) interacting with a quantized radiation field \cite{PhysRevB.84.195313}. Also, it allows one 
to investigate the $N-$photon Rabi splitting for two emitters having comparable coupling strengths \cite{Quesada}. 
As these calculations are meant to describe the outcome of optical measurements, the charge of each two-level subsystem is
assumed to be conserved.

More recently, experimental setups were extended to cavity-coupled double quantum emitters connected to 
source/drain particle reservoirs as key components of cavity-QED optoelectronics \cite{Kulkarni,Cottet_2017}. 
Deng {\it et al.} \cite{Deng} measured non-vanishing steady-state current correlations associated to a pair of 
distant graphene double QDs interacting with a microwave nanoresonator. The reflection amplitude of the latter displays a dip 
structure that was well fitted by the TC model and therefore proved the existence of non-local interaction through 
the microwave signal. In another work  the detection of electron-phonon interactions relied on transport measurements 
for a double-quantum dot defined in a suspended cavity-coupled InAs nanowire \cite{PhysRevLett.120.097701}.
 
Let us recall here that electrostatically coupled quantum wires and parallel quantum dots (QDs) display 
Coulomb drag and charge sensing effects at nanoscale which were experimentally observed \cite{Laroche631,Shinkai_2009,Bischoff} 
and extensively studied from the theoretical point of view 
\cite{PhysRevLett.104.076801,APJauho-drag,VMG,RevModPhys.88.025003}. 
Of course, this capacitive coupling becomes ineffective as the distance between the two systems increases. 
It would also seem that in the absence of optical or microwave probe signals the Tavis-Cummings coupling cannot be turned on.  

\begin{figure}[tbhp!]
 \includegraphics[angle=0,width=0.3\textwidth]{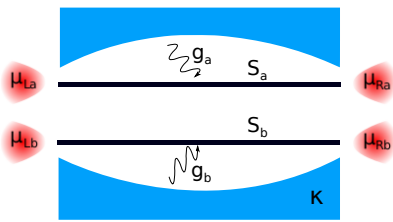}
 \caption{(Color online) Schematic view of two 1D nanowires $S_a$ and $S_b$ embedded in a single-mode cavity and individually
coupled to source and drain leads. $\mu_{is}$ is the chemical potential of the lead $i$ ($i=L,R$) attached to subsystem $s$ ($s=a,b$) 
and $g_s$ is the strength of the electron-photon interaction. The cavity losses are described by the parameter $\kappa$. 
The electron tunneling and the mutual Coulomb interaction between $S_a$ and $S_b$ are negligible.}
 \label{fig1}
 \end{figure}

We hereby theoretically show that the photon exchange between distant mesoscopic conductors embedded 
in a microcavity can be activated by electronic transport. Let us consider two parallel nanowires, each one
connected to a pair of source and drain particle reservoirs and embedded in a microcavity (see the sketch in Fig.\ \ref{fig1}).
The subsystems $S_a$ and $S_b$ accommodate electrons with both spin orientations and could be also quantum dots or carbon nanotubes. 
The electron tunneling and the mutual Coulomb interaction between $S_a$ and $S_b$ are negligible.

Clearly, electrons tunneling from a source reservoir may relax before tunneling out to the drain reservoir, 
while the emitted photons interact with both subsystems. Then, in analogy with classical electrodynamics, 
one can look for antenna-like coupling in which the current established in one system generates an 
electromagnetic field (i.e.\ photons) which changes the quantum state of the other system. 
We confirm this idea by calculating the transient and steady-state currents of the double-emitter cavity system
 for two transport settings (see Section III).  

On the theoretical side there are few works on transport trough multiple quantum emitters. Schachenmayer {\it et al.}
\cite{Schachenmayer,PhysRevB.97.205303} emphasized that the presence of the cavity enhances the current through 1D chains 
embedded in a single-mode cavity. 
Long distance coupling of resonant exchange qubits in the presence of the capacitive
coupling to a transmission line has been studied  by Russ and Burkhardt \cite{Russ}. However, to our best knowledge, time-dependent
transport calculations for distant parallel quantum emitters coupled by photons are not yet available.

The paper is organized as follows. The many-body Tavis-Cummings model, the structure of its dressed states and the 
non-markovian transport formalism  are presented in Section II. The numerical results are discussed in Section III. 
Conclusions are left to Section IV.

\section{Formalism}

\subsection{The Tavis-Cummings Hamiltonian of the double system}

We shall now consider the Tavis-Cummings model for our many-body systems. The Hamiltonian $H_s$ of each subsystem contains 
a non-interacting single-particle term and a two-particle Coulomb interaction within each subsystem (here $\sigma=\uparrow,\downarrow$ 
is the spin index and  $s=a,b$):
\begin{eqnarray}\label{H_s}
H_s&=&\sum_{i,\sigma}\varepsilon_{is}c_{i\sigma}^{\dagger}c_{i\sigma}+
\frac{1}{2}\sum_{\sigma,\sigma'}\sum_{i,j,k,l}v^{(s)}_{ijkl}c^{\dagger}_{i\sigma}c^{\dagger}_{j\sigma'}c_{l\sigma'}c_{k\sigma}.
\end{eqnarray}
The creation/annihilation operators $c_{i\sigma}^{\dagger}/c_{i\sigma}$ are associated to spin-dependent single-particle states 
$\psi^{(s)}_{i\sigma}$ of each subsytem. The eigenvalues $\varepsilon_{is}$ are obtained by diagonalizing the
single-particle Hamiltonian of the non-interacting double system. The wavefunctions $\psi^{(s)}_{i\sigma}$ inherit the size 
and geometry of the 1D nanowire. The matrix elements of the Coulomb interaction $v^{(s)}_{ijkl}$ are 
then calculated in terms of single-particle states $\psi^{(s)}_{i\sigma}$ :
\begin{equation}\label{C-mat}
v_{ijkl}^{(s)}=\sum_{\alpha,\beta}\overline{\psi^{(s)}_{i\sigma}(\alpha)}\psi^{(s)}_{k\sigma}(\alpha)u(\alpha-\beta)
\overline{\psi^{(s)}_{j\sigma'}(\beta)}\psi^{(s)}_{l\sigma'}(\beta),
\end{equation}
where $\alpha, \beta$ are sites describing the subsystem $S_s$, $\overline{\psi^{(s)}_{i\sigma}}$ is the complex conjugate of the 
single-particle wavefunction and $u(\alpha-\beta)$ is the Coulomb potential. A small screening constant is added in the Coulomb 
kernel in order to avoid on-site singularities.

 The interacting many-body states (MBSs) $|\nu\rangle$ 
and the associated energies $E_{\nu}$ of the double system are defined as: 
\begin{equation}\label{MBS-P}
(H_a+H_b)|\nu\rangle=E_{\nu}|\nu\rangle.
\end{equation}
Embedding this parallel structure in a single-mode cavity of frequency $\omega$  results in a hybrid system described by:
\begin{equation}\label{TCH}
H_S=H_a+H_b+\hbar\omega {\hat a}^{\dagger}{\hat a}+\sum_{s=a,b}V_s:=H_S^{(0)}+V_{{\rm el-ph}},
\end{equation}
where $V_s$ stands for the optical coupling between electrons in subsystem $s$ and the cavity photons:
\begin{equation}\label{V_JC}
V_s=\sum_{i,j\in S_s}\sum_{\sigma}\hbar g^{(s)}_{ij}c_{i\sigma}^{\dagger}c_{j\sigma}({\hat a}^{\dagger}+{\hat a}).
\end{equation}
Note that the Hamiltonian $H_S$ in Eq.\ (\ref{TCH}) is more general than the Tavis-Cummings Hamiltonian encountered in quantum optics, 
as it acts on a many-body configuration space which includes the empty states of each subsystem and the spin degree of 
freedom. The optical selection rules are embodied in the constants $g^{(s)}_{ij}$, and in particular the 
spin $\sigma$ is conserved. $a^{\dagger}$ denotes the photon creation operator and 
$\hbar\omega {\hat a}^{\dagger}{\hat a}|N\rangle=N\hbar\omega|N\rangle$ where $|N\rangle$ is the $N$-photon Fock state. 
The coupling constants are calculated as
\begin{equation}\label{g_ij}
g^{(s)}_{ij}=\frac{e}{m_0}\sqrt{\frac{\hbar}{2\epsilon\omega V}}
\langle\psi^{(s)}_{i\sigma}|{\bf e}\cdot {\bf p}|\psi^{(s)}_{j\sigma} \rangle,
\end{equation}
where ${\bf p}$ is the momentum operator, ${\bf e}$ is the polarization vector, $\epsilon$ is the dielectric constant and $V$ 
is the volume of the cavity. The matrix elements $g_{ij}^{(s)}$ in Eq.\ (\ref{g_ij}) are
calculated numerically by discretizing the momentum operator and using its action on the site-dependent
single-particle wavefunctions. 

In the rotating wave approximation Eq.\ (\ref{V_JC}) counts only terms for which $\varepsilon_{js}<\varepsilon_{is}$ and 
$V_s$ reduces to the well known Jaynes-Cummings (JC) optical coupling. We shall use the simplified notation 
$g_{12}^{(s)}=\overline{g_{21}^{(s)}}:=g_s$. Moreover, for identical subsystems 
one has $g_a=g_b=g_0$.
  
\subsection{Energy spectrum and dressed states}
 
In order to capture the main physics of the open hybrid system  we shall adopt here a simple lattice model. A more accurate 
description of the cavity-coupled system requires a continuous model in spatial coordinates which was implemented 
in previous work \cite{Gudmundsson12:1109.4728,Jonasson2011:01,Gudmundsson:2013.305}.

Let $\varepsilon_{1s,2s}$ be the lowest spin-degenerate single-particle energies of subsystem $s$, ordered such that
$\varepsilon_{1s}<\varepsilon_{2s}$ (i.e. $i=1,2$). In the absence of both Coulomb interaction and electron-photon coupling
the many-body states of the double system are written in the occupation number basis associated to the single-particle states
$\psi^{(s)}_{i\sigma}$. The occupation of such a state is specified by the spin $\sigma_{is}$. For example the state
$|\uparrow_{1a}\downarrow_{2b}\rangle$ contains one electron on each subsystem, occupying the levels $\varepsilon_{1a}$ and
$\varepsilon_{2b}$ and having the indicated spin orientations.

We start by diagonalizing the interacting many-body Hamiltonian $H_a+H_b$ of the double system
on a reduced Fock space comprising all 256 non-interacting many-body configurations containing up to 4 electrons
which are allowed to occupy the single-particle levels $\varepsilon_{1s,2s}$ of each subsystem $S_s$. For suitable
values of the bias voltage applied on each nanowire the resulting interacting many-body states $|\nu\rangle$
provide a reliable basis for transport calculations. This choice reduces considerably the numerical cost of the 
time-dependent transport calculations but also captures the optical processes involving only pairs of spin-degenerate 
single-particle lowest-energy states of each subsystem. The exact diagonalization method is not pertubative and therefore 
a suitable 'small' parameter does not present itself. Nonetheless, error estimates for interacting quantum dots have been 
calculated (see e.g.\ the work of Kvaal, \cite{PhysRevB.80.045321} and the paper by Jeszenszki {\it et al}.\ on 1D quantum gases 
with contact interactions \cite{PhysRevA.98.053627}). For the parameters selected in our calculations the relevant 
interacting many-body states are numerically stable when computed by diagonalizing the Hamiltonian on several truncated subspaces.
Moreover, since the electron-photon coupling strength $g_0\ll\omega$ the accuracy of the dressed states is even
higher. The convergence of the exact diagonalization method for circuit quantum electrodynamics was thoroughly
investigated in a previous publication \cite{PhysRevE.86.046701}. 

Since there is no tunneling between $S_a$ and $S_b$ the electronic occupations $n_s^{(\nu)}$ of each subsystem $s$ for a given 
MB configuration $|\nu\rangle$ are good quantum numbers: 
\begin{equation}\label{n_s}
n_s^{(\nu)}:=\sum_{i\in S_s,\sigma}\langle\nu|c^{\dagger}_{i\sigma}c_{i\sigma}|\nu\rangle=\sum_i\,n_{is}^{(\nu)}. 
\end{equation}

Then the eigenstates $|\nu,N\rangle$ of the disjointed Hamiltonian $H_S^{(0)}$ (see Eq.\ (\ref{TCH})) can be organized in
orthogonal subspaces labeled by particle and photon numbers $(n_a,n_b;N)$. For further use let us briefly describe
these subspaces. The $N$-photon `empty' states (i.e.\ without electrons) of the subspace $(0,0;N)$ are denoted
by $|0,N\rangle$. Next, one has four single-electron states $|\sigma_{is},N\rangle$ for each subsystem, leading to the subspaces
$(1,0;N)$ and $(0,1;N)$. The subspace $(1,1;N)$ comprises 16 two-particle states which, when mixed by the electron-photon
coupling generate the Tavis-Cummings dressed states (see below). Note that because the Coulomb interaction  between the
two subsystems is neglected the configurations belonging to this subspace can be simply written as $|\sigma_{ia}\sigma_{jb},N\rangle$.

The internal Coulomb interaction effectively shows up in the subspaces $(2,0;N)$ and $(0,2;N)$. The degenerate
two-particle ground states are $|G^{(s)},N\rangle:=|\uparrow_{1s}\downarrow_{1s},N\rangle$, while the antiparallel and
parallel triplet configurations are denoted by $|T^{(s)}_0,N\rangle$, $|T^{(s)}_1,N\rangle$. Finally,
$|S^{(s)},N\rangle$ stands for the singlet configurations. More complicated configurations can be constructed in a similar
way.

 Let ${\cal E}^{(0)}_{\nu,N}=E_{\nu}+N\hbar\omega$ be the energy of the `free' state defined by 
$H_S^{(0)}|\nu,N\rangle={\cal E}^{(0)}_{\nu,N}|\nu,N\rangle$. The fully interacting electron-photon Hamiltonian $H_S$ is then 
diagonalized w.r.t.\ the  basis $\{|\nu,N\rangle\}$ of the disjointed systems. Its eigenfunctions and eigenvalues are denoted by 
$|\varphi_p\rangle$ and ${\cal E}_p$ such that
\begin{equation}\label{dressed}
H_S|\varphi_p\rangle={\cal E}_p|\varphi_p\rangle.
\end{equation}
 Here $p$ is a set of relevant quantum numbers (see below). 
In the transport calculations the number of photons is truncated to $N_{{\rm ph}}$, that is we allow at most 
$N_{{\rm ph}}+1$ Fock states to assist the transport.
The electron-photon coupling mixes the `free' states $|\nu,N\rangle$ but one finds that, besides the electronic occupations 
$n_s$ on each subsystem, the excitation number $x=\sum_s\,n_{2s}+N$ is 
also conserved. Here $n_{2s}$  is the occupation of the excited single-particle level $\varepsilon_{2s}$ of $S_s$. 

Up to spin-dependent quantum numbers, the fully interacting states are also organized in several subspaces
 described by the excitation number $x$ and partial occupations $n_a$, $n_b$. 
Obviously the `empty' states are stable against the electron-photon coupling and one has $|\varphi_{N,0}\rangle:=|0,N\rangle$. 
For the single-particle sector ($n_a+n_b=1$) we get spin degenerate (optically active) dressed states $|\varphi^{\pm}_{N,\sigma_s}\rangle$ 
for each two-level system  and some dark states:
\begin{eqnarray}\label{dressed-JC1}
|\varphi^{\pm}_{N,\sigma_s}\rangle&=&\frac{1}{\sqrt 2}(|\sigma_{1s},N+1\rangle \pm |\sigma_{2s},N\rangle),\\
\label{dressed-JC2}
|\varphi_{0,\sigma_{1s}}\rangle &=&|\sigma_{1s},0\rangle,\quad  |\varphi_{N_{{\rm ph}},\sigma_{2s}}\rangle =|\sigma_{2s},N_{{\rm ph}}\rangle.
\end{eqnarray}
The excited state $ |\varphi_{N_{{\rm ph}},\sigma_{2s}}\rangle$ given by Eq.\ (\ref{dressed-JC2}) cannot emit another photon because of
truncation w.r.t.\ the Fock states. The energies of the dressed states $|\varphi^{\pm}_{N,\sigma_s}\rangle$ at 
resonance, that is when $\varepsilon_{2s}-\varepsilon_{1s}=\hbar\omega$ ($N\geq 0$), are:
\begin{equation}\label{En-JC1}
{\cal E}^{\pm}_{N,\sigma_s}=\varepsilon_{2s}+N\hbar\omega\pm \frac{\hbar\Omega_N}{2},
\end{equation}
where $\Omega_N=2 g_{0}\sqrt{N+1}$ is the well known $N$-photon Rabi frequency of the two-level JC model. 
The electron-photon interaction also affects the two-particle sector ($n_a+n_b=2$, $n_a=n_b=1$). Let us introduce first 
the ground ($G$), doubly-excited ($X$), triplet ($T$) and singlet ($S$) spin-dependent  Dicke states:
\begin{eqnarray}\label{G_ssp}
|G_{\sigma\sigma'},N\rangle&=&|\sigma_{1a}\sigma'_{1b},N\rangle,\\\label{X_ssp}
|X_{\sigma\sigma'},N\rangle&=&|\sigma_{2a}\sigma'_{2b},N\rangle,\\\label{T_ssp}
|T_{\sigma\sigma'},N\rangle&=&\frac{1}{\sqrt 2}\left (|\sigma_{2a}\sigma'_{1b},N\rangle+|\sigma_{1a}\sigma'_{2b},N\rangle \right ),\\\label{S_ssp}
|S_{\sigma\sigma'},N\rangle&=&\frac{1}{\sqrt 2}\left (|\sigma_{2a}\sigma'_{1b},N\rangle-|\sigma_{1a}\sigma'_{2b},N\rangle \right ).
\end{eqnarray}
For identical emitters it can be shown that at resonance the two-particle 
dressed states have the following structure (for the simplicity of writing we omitted the spin indices of the 
two-particle configurations): 
\begin{eqnarray}\nonumber
|\varphi^{(1)}_{N,\sigma\sigma'}\rangle&=&\sqrt{\frac{N}{2N-1}}|X,N-2\rangle-\sqrt{\frac{N-1}{2N-1}}|G,N\rangle ,  \\
\nonumber
|\varphi^{(2,3)}_{N,\sigma\sigma'}\rangle&=&\sqrt{\frac{N}{4N-2}}|G,N\rangle\pm\frac{1}{\sqrt{2}}|T,N-1\rangle \\
\label{dressed-TC23}
&+&\sqrt{\frac{N-1}{4N-2}}|X,N-2\rangle, \\\nonumber
\varphi^{(4)}_{N,\sigma\sigma'}\rangle&=&|S,N-1\rangle .
\end{eqnarray}
The above expressions generalize the spinless case discussed by Quesada \cite{Quesada}. Note these expressions 
hold as long as the Coulomb interaction between the two subsystem is negligible. 
One infers that the states $|\varphi^{(\alpha)}_{N,\sigma\sigma'}\rangle$
with $\alpha=2,3,4$ exist only for $N\geq 1$ while 
$|\varphi^{(1)}_{N,\sigma\sigma'}\rangle $ is not defined for $N=1$ and 
$|\varphi^{(1)}_{0,\sigma\sigma'}\rangle=|G_{\sigma\sigma'},0\rangle $. For a non-vanishing excitation number $N$ one gets 
a subspace of Tavis-Cummings dressed states $\{|\varphi^{(\alpha)}_{N,\sigma\sigma'}\rangle\}$.  If the cavity mode is 
slightly detuned from resonance the eigenvalues ${\cal E}^{(\alpha)}_{N,\sigma\sigma'}$ are still four-fold degenerate w.r.t.\ 
the spin indices and the coefficients of the `free' states can only be obtained by numerical diagonalization. The structure of the 
fully interacting states is however not affected (i.e. the electron-photon coupling mixes the same `free' states). 

As for the corresponding energies one finds that at resonance:
\begin{eqnarray}\label{En-TC14}
{\cal E}^{(1)}_{N,\sigma\sigma'}&=&{\cal E}^{(4)}_{N,\sigma\sigma'}=\varepsilon_{1a}+\varepsilon_{1b}+N\hbar\omega\\\label{En-TC23} 
{\cal E}^{(2,3)}_{N,\sigma\sigma'}&=&\varepsilon_{1a}+\varepsilon_{1b}\pm \hbar g_0\sqrt{4N-2}+N\hbar\omega. 
\end{eqnarray}
From Eqs.\ (\ref{En-TC14}) and (\ref{En-TC23}) we infer that within the Tavis-Cummings $N$-excitation subspace the dynamics is controlled by 
two Rabi frequencies $\tilde\Omega_N=2g_0\sqrt{4N-2}$ and $\tilde\Omega_N/2$ associated to the two spectral gaps 
${\cal E}^{(2)}_{N,\sigma\sigma'}-{\cal E}^{(3)}_{N,\sigma\sigma'}$ and ${\cal E}^{(1)}_{N,\sigma\sigma'}-{\cal E}^{(2)}_{N,\sigma\sigma'}$. In fact by integrating numerically
 the von Neumann equation of the closed hybrid system described by $H_S$ (see Eq.\ (\ref{TCH})) one finds that the
populations of the optically active `free states' and the mean photon number oscillate with periods associated to the above frequencies.

\subsection{Generalized master equation method}

We set our transport problem in the partitioning approach \cite{Caroli_1971} by assuming that at some time instants $t_{Ls}$ and $t_{Rs}$ 
the subsystem $S_s$ is smoothly coupled to left (L) and right (R) particle reservoirs having chemical potentials $\mu_{Ls},\mu_{Rs}$ 
(see the sketch in Fig.\ \ref{fig1}). 
The reservoirs are modeled as semiinfinite tight-binding chains supplying electrons with both spin orientations. 
The Hamiltonian of the open system is written as 
\begin{equation}
H(t)=H_S+H_L+H_T(t),
\end{equation}
where $H_L$ is associated to the four leads and $H_T$ is the lead-sample tunneling term containing time-dependent smooth switching 
functions $\chi_l$ (here $l=L_a,R_a,L_b,R_b$):
\begin{eqnarray}\label{H_leads}
H_L&=&\sum_l\sum_{\sigma}\int dk\,\varepsilon_{k\sigma l}c^{\dagger}_{k\sigma l}c_{k\sigma l}\\\label{Htunnel}
H_T(t)&=&\sum_{s,l}\sum_{i\in S_s,\sigma}\int dk \chi^l(t)(T^{(ls)}_{ki}c^{\dagger}_{k\sigma l}c_{i\sigma}+h.c),
\end{eqnarray}
where $c^{\dagger}_{k\sigma l}$ is the creation operator on lead and $k$ is the electronic momentum. For simplicity we impose 
spin conservation in the tunneling region such that the coupling parameter $T^{(ls)}_{ki}$ is spin-independent. 
In the present model we take into account the dependence of the tunneling coefficient on the single-particle 
wavefunctions \cite{Moldoveanu09:073019}, that is $T^{(ls)}_{ki}=V_{l,s}{\phi}^{l*}_k(0_l)\psi^{(s)}_{i\sigma}(n_l)$ where $0_l$ 
is the site of the lead $l$ which couples to the contact site $n_l$ of the corresponding subsystem. $V_{l,s}$ is a constant input 
parameter. The four-lead geometry shown in Fig.\ \ref{fig1} corresponds to non-vanishing parameters $V_{L_s,s}:=V_{Ls}$ and $V_{R_s,s}:=V_{Rs}$.  
We tune $V_{l_s}$ such that the values of the tunneling rates $\Gamma^{(Ls)}_{ki}=|T^{(ls)}_{ki}|^2$ are around few $\mu$eV.
The spectrum of the semiinfinite leads is $\varepsilon_{k\sigma l}=2t_L\cos k$, where $t_L$ denotes the common hopping energy on 
the leads.

Using the Nakajima-Zwanzig projection technique one obtains an equation for the reduced density operator (RDO) of the hybrid system
$\rho(t)={\rm Tr}_L\{W\}$, where $W$ is the full density operator of the coupled system and ${\rm Tr}_L$ is the trace over 
the leads' degrees of freedom:
\begin{eqnarray}\nonumber
&&{\dot\rho}(t)=-\frac{i}{\hbar}[H_S,\rho(t)]\\\nonumber
&&-\frac{1}{\hbar^2}{\rm Tr}_L\left\lbrace\left[ H_{\rm T}(t),\int_{t_0}^t ds U_{t-s} \left[H_{\rm T}(s),\rho(s)\rho_L\right ]
U^{\dagger}_{t-s}\right ]\right\rbrace \\\label{GME1}
&&-\frac{\kappa}{2}\left (a^{\dagger}a\rho(t)+\rho(t) a^{\dagger} a-2a\rho(t) a^{\dagger} \right ).
\end{eqnarray}
In Eq.\ (\ref{GME1}) $U_t=e^{-it(H_S+H_L)/\hbar}$ is the unitary evolution of the disconnected systems and $\rho_L$ is the 
equilibrium density operator of the leads. The third line defines a Lindblad-type operator which takes into account cavity losses.

Note that in this basis the unitary evolution $U_t$ is easily handled as it
 becomes a diagonal matrix. The matrix form of GME leads us to consider transitions between pairs of states $\{\varphi_p,\varphi_{p'} \}$ 
\cite{DMG}. As an example, the generalized transition matrix element:
\begin{equation}\label{A-tunn}  
{\cal A}^{(l)}_{pp'}(k)=\sum_s\sum_{i\in S_s,\sigma}T^{(ls)}_{ki}\langle\varphi_p|c^{\dagger}_{i\sigma}|\varphi_{p'}\rangle 
f_l(\varepsilon_{k\sigma l}),
\end{equation}
 captures the tunneling processes of electrons from the $l$-th lead to all single-particle levels $i$ of the parallel structure.
The tunneling selection rules can be obtained by considering the non-vanishing 
matrix elements of the creation operator w.r.t.\ the basis $\{\varphi_p\}$. 
In the steady-state regime one expects to recover the Born-Markov approximation such that the 
tunneling is controlled by Fermi-Dirac weights $f_l({\cal E}_p-{\cal E}_{p'})$. This points out that the energy ${\cal E}_p-{\cal E}_{p'}$ 
required to add an extra electron to some initial configuration $|\varphi_{p'}\rangle$ of the parallel structure must be below the chemical 
potential of the $l$-th lead in order to allow the tunneling process leading to the final state $|\varphi_{p}\rangle$. 

As an example, the simplest transition between the single-particle groundstate $|\varphi_{0,\sigma_{1s}}\rangle$ to the 
dark two-particle groundstate $|G_s\rangle=|\uparrow_{1s}\downarrow_{1s}\rangle$ 
requires the energy $\mu_{G_s}={\cal E}_{{G_s}}-{\cal E}_{0,\sigma_{1s}}=\varepsilon_{1s}+U$, where $U$ is a few meV shift due to the 
internal Coulomb interaction. If $\mu_{Ls,Rs}<\mu_{G_s}$ this tunneling process is suppressed and the double occupancy of the 
subsystem $s$ is excluded. 

The GME is solved numerically as an integro-differential  system of coupled equations for the matrix elements of $\rho$ w.r.t\ the
basis of dressed states $\{|\varphi_p\rangle\}$.
Once the RDO is calculated the mean value of the total charge operator $Q_S=\sum_{s,i,\sigma}c^{\dagger}_{i\sigma}c_{i\sigma}$ 
is calculated as $\langle Q_S(t)\rangle={\rm Tr}_{\cal F}\{Q_S\rho(t)\}$, the trace being performed on the Fock space ${\cal F}$ 
made by the eigenstates $|\varphi_p\rangle$ of the hybrid system. The left and right transient currents $J_{Ls}, J_{R_s}$ are 
identified from the continuity equation:
\begin{equation}\label{J-current}
\frac{d}{dt}\langle Q_S\rangle={\rm Tr}_{\cal F}\{Q_S{\dot\rho}(t)\}=\sum_s\, \left (J_{Ls}(t)-J_{Rs}(t)\right ).
\end{equation}
By using the cyclicity of the trace and fact that $Q_S$ commutes with the bosonic operators one easily finds that loss term 
in the GME does not contribute directly to the currents. Nonetheless, it affects the matrix elements of the RDO which are fully 
contained in the dissipative term due to the leads. The photon number is given by:
\begin{equation}\label{N_ph}
{\cal N}(t)={\rm Tr}_{{\cal F}}\{\rho(t){\hat a}^{\dagger}{\hat a} \}.
\end{equation}
The average charge occupation of the single-particle levels $\varepsilon_{is}$ is given by 
$q_{is}(t)=e\sum_{\sigma}{\rm Tr}_{\cal F}\{\rho(t)c^{\dagger}_{i\sigma}c_{i\sigma}\}$ (here $e$ denotes the electron charge). 
It is also useful to introduce the total populations of the Jaynes-Cummings ($n_a+n_b=1$) and Tavis-Cummings ($n_a=n_b=1$) $N$-photon manifolds:
\begin{eqnarray}\label{P_JC}
P_{JC,N}^{\pm}(t)&=&\sum_{s,\sigma}\langle\varphi^{\pm}_{N,\sigma_s}|\rho(t)|\varphi^{\pm}_{N,\sigma_s}\rangle,\\\label{P_TC}
P_{TC,N}^{(\alpha)}(t)&=&\sum_{s,\sigma}\langle\varphi^{(\alpha)}_{N,\sigma_s}|\rho(t)|\varphi^{(\alpha)}_{N,\sigma_s}\rangle .
\end{eqnarray}
The total populations of JC and TC states are then easy to calculate (we omit the time-dependence for the simplicity of writing):
\begin{eqnarray}\label{P_TC_total}
P_{JC,N}=\sum_{\lambda=\pm}P_{JC,N}^{(\lambda)},\quad P_{TC,N}=\sum_{\alpha}P_{TC,N}^{(\alpha)}
\end{eqnarray}

In the present work we use a generalized master equation approach which provides the dynamics of the many-body 
configurations in the presence of sequential tunneling processes. Other methods allow the calculations of full 
counting statistics (FCS) for interacting systems \cite{PhysRevB.94.214308,PhysRevB.97.115109}. In particular, in 
the presence of driving voltages the stochastic path-integral approach of Altland {\it et al.} 
\cite{PhysRevLett.105.170601,PhysRevB.82.115323} predicts that massive fluctuations will exceed the average values. 

\section{Numerical results and discussion}

Our parallel structure comprises two identical 1D nanowires of 50 nm each, described as a lattice chain 
of $N_x=250$ sites. The lowest two single-particle energies are $\varepsilon_{1s}=41.25$ meV and $\varepsilon_{2s}=41.95$ meV.  
The numerical calculations were performed by taking into account up to two photons in the cavity and the thermal energy 
$k_BT=50$\ mK. However, at the end of this Section we provide and discuss results obtained when the number of Fock states included
in the calculations increases. Moreover, we neglected the cavity losses as for $\kappa\ll g_0$ we obtained similar results. 
The hopping energy on the leads $t_L=1$meV.    

 The interplay of sequential tunneling and photon emission/absorption processes leads to a complicated dynamics of 
the hybrid system. However one expects that for weak coupling to the leads some features of the unitary dynamics $U_t=e^{-itH_S/\hbar}$ 
of the closed system could still be present in the transport properties. In particular, the time-dependent 
charge occupations on each subsystems will be shown to exhibit periodic JC or TC Rabi oscillations.   

We set the chemical potentials of the source leads $\mu_{La}$ and $\mu_{Lb}$ above the single-particle levels $\varepsilon_{is}$ 
but well below the energies of the interacting two-particle configurations of the type $(n_a=2, n_b=0)$ or $(n_a=0, n_b=2)$. Then 
the internal Coulomb interaction prevents the population of states with more than one electron on each system. These states can be 
safely disregarded in the time-dependent transport calculations and the GME is solved within a truncated subspace made of 25 
electronic MBSs for the double system and the Fock states containing up to two photons. In this setup electrons tunnel through 
the system only via configurations with one electron and the relevant optical transitions only involve the lowest single-particle levels 
$\varepsilon_{1s}$ and  $\varepsilon_{2s}$. 

\subsection{Removal of the Coulomb blockade}

\begin{figure}[t]
\includegraphics[angle=0,width=0.475\textwidth]{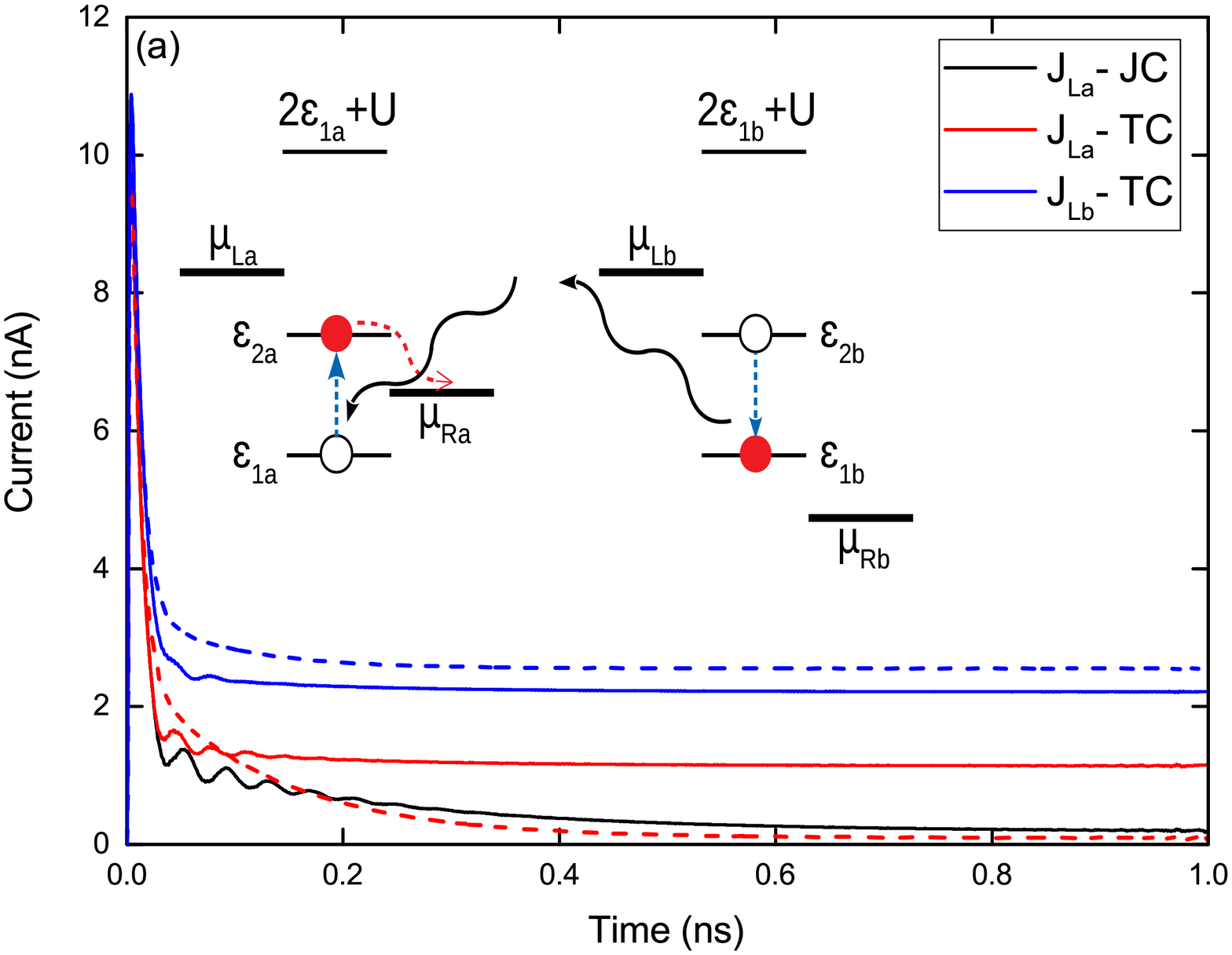}
\includegraphics[angle=0,width=0.475\textwidth]{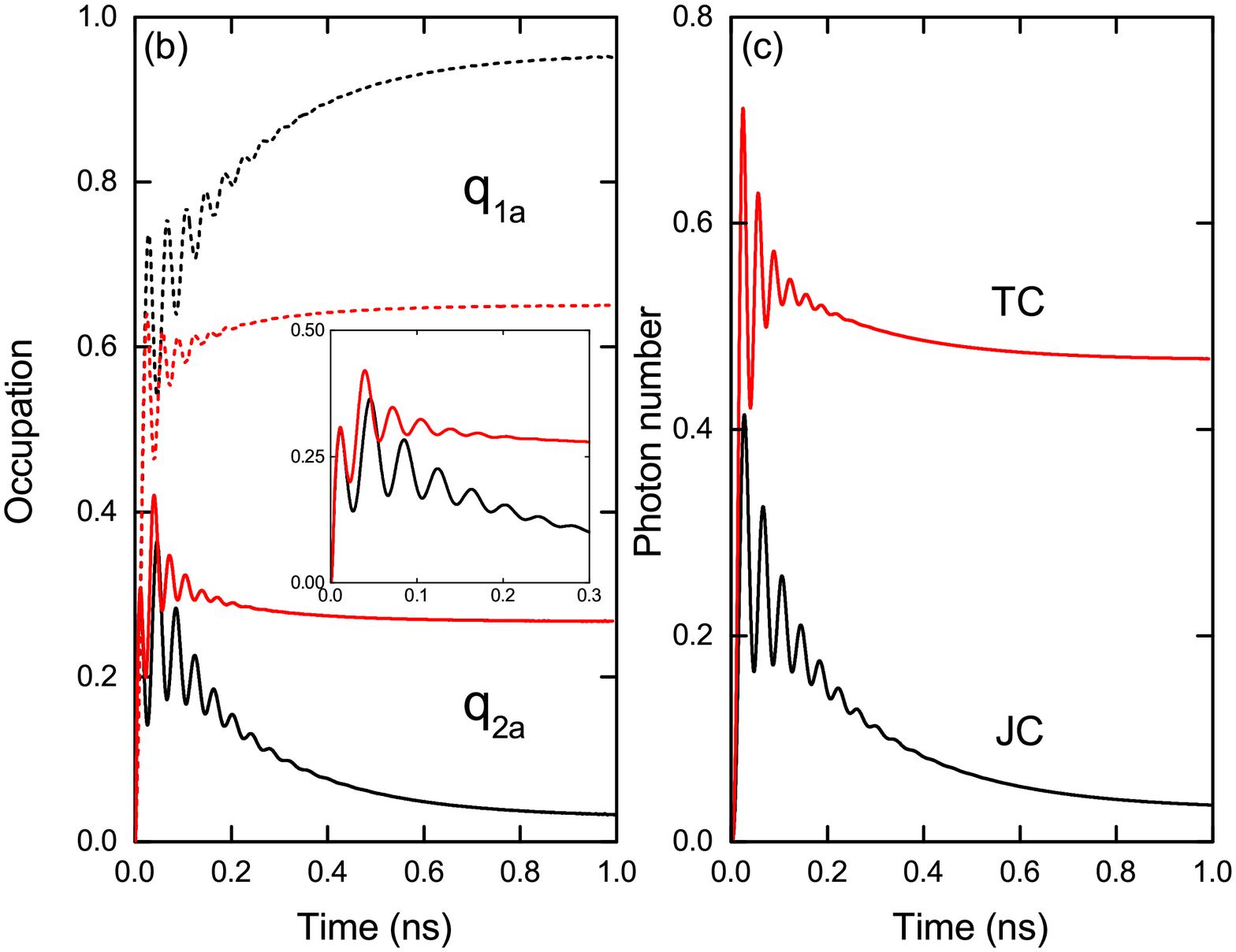}
 \caption{(Color online) a) The current $J_{La}$ (black solid line) vanishes in the steady-state if the lower subsystem is disconnected. 
A current passing $J_{Lb}$ on the lower subsystem $S_b$ (blue solid line) removes the Coulomb blockade as suggested in the inset and
a non-vanishing steady-state value of  $J_{La}$ is noticed (red solid line). The dashed lines represent the transient currents 
in the absence of the electron-photon coupling. Inset: A sketch of the low-energy levels of each subsystem 
and of the relevant processes in the removal of the Coulomb blockade: photon emission/absorption - black solid wavy line, electron relaxation/excitation - vertical dashed blue line and tunneling to the leads - dashed red line. 
b) The charge occupations on the optically active levels of $S_a$ for the Jaynes-Cummings (dashed line) and Tavis-Cummings 
(solid line) transport configurations. Inset: $q_{2a}$ in the transient regime. c) The mean photon number for JC and TC regimes.  
Other parameters: 
$\mu_{La}=\mu_{Lb}=45$\ meV, $\mu_{Ra}=41.5$\ meV, $\mu_{Rb}=35$ meV. The electron-photon coupling strength $g_0=53$ $\mu$eV, $\kappa=0$.}
 \label{fig2}
 \end{figure}

The first transport setup we considered reveals the switching from the Jaynes-Cummings dynamics of a single-subsystem to the 
Tavis-Cummings dynamics of the photon-coupled double system. The chemical potential $\mu_{R_a}$ is chosen such that the  
single-particle level $\varepsilon_{2a}$ lies well within the bias window $\mu_{La}-\mu_{Ra}$ and the lowest one $\varepsilon_{1a}$ is 
below $\mu_{Ra}$ (see the inset in Fig.\ \ref{fig2}(a). The fact that each subsystem accomodate only up to one electron is 
also suggested by indicating the energy of the ground two-particle states $2\varepsilon_{1s}+U$, where $U$ denotes the upward 
shift due to the internal Coulomb interaction. The chemical potential $\mu_{Rb}$ is set such that both single-particle 
levels of $S_b$ are within the bias window. We use equal couplings to the leads, $V_{Ls}=V_{Rs}$.

We calculated the current passing through the upper nanowire ($S_a$) while keeping the lower system $S_b$ disconnected from leads. 
The initial state is $|00,0\rangle$, that is each nanowire is empty and there are no photons in the cavity. In this case 
 the only optically active states belong to the JC subspace $|\varphi^{\pm}_{N,\sigma_a}\rangle$ since $n_a=1$ and $n_b=0$. 
Next, we repeat the calculations for the same initial state except that now {\it both} systems are simultaneously coupled 
to the leads in order to populate two-electron Tavis-Cummings dressed-states. 

Fig.\ \ref{fig2}(a) presents the transient currents corresponding to the JC and TC transport configurations.
In the absence of the second subsystem  the steady-state current vanishes due to the Coulomb blockade and the lowest level 
$\varepsilon_{1a}$ is nearly filled. This can be seen in the average electronic occupation of the ground ($q_{1a}$) and excited ($q_{2a}$) 
single-particle energy levels in Fig.\ \ref{fig2}(b)). The charge occupations display few Rabi oscillations on their way to the 
steady-state. Similar oscillations of the transient current were reported for a continuous model \cite{doi:10.1021/acsphotonics.5b00115}. 
Here we find that in the JC regime the oscillation period is $T_0=39$\,ps, which corresponds to the Rabi frequency $\Omega_{0}$. 
The average photon number also vanishes (see Fig.\ \ref{fig2}(c)) because the steady-state configuration is an equal weight combination 
of ground states $|G_{\sigma\sigma'},0\rangle$ (not shown).

 In Fig.\ \ref{fig2}(a) we also show the transient currents $J_{La}$ and $J_{Lb}$ in the absence of the electron-photon
coupling (see the blue and red dashed lines). As expected, the current through the system $S_a$ vanishes due to the
Coulomb blockade and no Rabi oscillations are noticed. In contrast, since the bias window on $S_b$ allows tunneling processes
even in the steady-state and the photon exchange with $S_a$ is no longer present, $J_{Lb}$ reaches a slightly larger value
in the stationary regime.

Note that for excitonic systems like self-assembled QDs the corresponding ground state is the fully occupied
valence band which can only be populated via electron relaxation from the conduction band followed by photon emission. In 
our system {\it both} `conduction' ($\varepsilon_{2s}$) and `valence' ($\varepsilon_{1s}$) levels can be fed from 
the source contacts. The direct tunneling to the lower single-particle states considerably hampers photon emission so 
we can restrict our numerical calculations to few Fock states only.   

In the TC setting the above picture changes considerably. Fig.\ \ref{fig2}(a) shows that the current $J_{La}$ does not 
vanish anymore in the steady-state, which means that the tunnelings from the excited level $\varepsilon_{2a}$ to the right 
contact are now allowed. This removal of the Coulomb blockade on subsystem $S_a$ proves the correlations due to the photon exchange 
between the two systems. The mechanism leading to the removal of the Coulomb blockade is also suggested in the inset of 
Fig.\ \ref{fig2}(a): i) With two levels within its bias window, subsystem $S_b$ generates photons even in the steady-state 
(see the corresponding mean photon number in Fig.\ 2(c)); ii) A photon emitted by $S_b$ excites electrons from the lowest level 
of $S_a$ to the higher level which in turn tunnels to the right lead and contributes to the transport. 

This scenario is confirmed by Fig.\ 2(b) from which one checks that the occupation $q_{2a}$ is now around 0.25 while the lowest level is not
fully occupied. We also notice that in the TC regime the Rabi oscillations have a different period (see the inset of Fig.\ 2(b)). 
This is expected because the JC and TC subspaces have different Rabi frequencies for the associated dynamics. 
We shall discuss this fact in the next subsection.    

The steady-state current $J_{Lb}$ through the lower subsystem is larger that $J_{La}$ due to the fact that both levels $\varepsilon_{ib}$ 
are within the bias window. Let us note that the position of the two levels w.r.t.\ bias window is crucial for the removal of the Coulomb
blockade. We have checked that by decreasing the bias window on $S_b$ (i.e. for $\mu_{Rb}=41.5$ meV) both systems are simultaneously 
blocked in the steady-state as their lowest levels are fully charged and below the bias window.

\subsection{All-electrical `reading' of a closed system}

The setup presented in the previous subsection captures a steady-state effect of the photonic coupling 
of the two electronic systems. To illustrate transient effects of the photon-mediated interaction between the 
two subsystems let us consider a transport setup in which the lower nanowire $S_b$ is again disconnected from 
leads but can still be correlated to the upper wire. We assume that subsystem $S_b$ is prepared in some initial state 
$|\nu_b\rangle$ {\it before} passing a current through the nearby subsystem $S_a$ at instant $t_0=0$. If the two 
systems exchange photons (i.e.\ when $n_b\neq 0$ and the resonant condition holds both for $S_a$ and $S_b$) the transport 
through the open system $S_a$ should depend on the dynamics in the closed system $S_b$. The chemical potential of the 
leads attached to $S_a$ are set such that both single-particle levels are within the bias window and photons are 
generated even in the steady-state regime. 

\begin{figure}[t]
 \includegraphics[angle=0,width=0.45\textwidth]{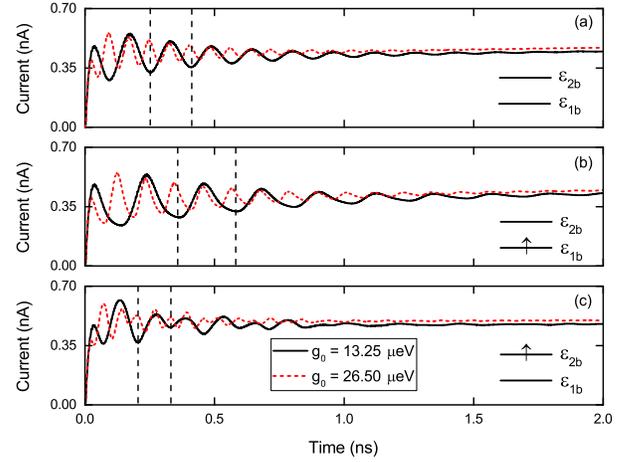}
 \caption{(Color online) The transient Rabi oscillations of the output current $J_{Ra}$ corresponding to different initial 
states of the closed system. a) $|0\rangle$, b) $|\uparrow_{1b}\rangle$ and c) $|\uparrow_{2b}\rangle$.
The vertical dashed lines mark the Rabi oscillation period. The three initial configurations are also shown in the inset.
Other parameters: $\mu_{La}=\mu_{Lb}=45$ meV, $\mu_{Ra}=\mu_{Rb}=35$ meV, $\kappa=0$.}
 \label{fig3}
 \end{figure}

In Fig.\ \ref{fig3} we show the transient currents $J_{Ra}$ for three initial configurations $|\nu_b\rangle$ of the closed system 
specified by the 
spin-dependent occupation numbers of the single-particle states in $S_b$, i.e. $|0\rangle$, $|\uparrow_{1b}\rangle$ and 
$|\uparrow_{2b}\rangle$.  Although for simplicity we considered only a spin up electron occupying the subsystem $S_b$ 
a mixture of both spin orientations could be initialized as well without changing the results. Each figure contains 
results for two values of the electron-photon coupling strength, $g_0=13.25$ $\mu$eV and $g_0=26.5$ $\mu$eV. Note that the Rabi 
period increases as $g_0$ decreases. Also, to avoid a fast damping of the Rabi oscillations we reduced the coupling to the leads.  

One notices at once that the output transient current develops oscillations whose period depends on the initial state 
of subsystem $S_b$. In fact, the three periods in Fig.\ \ref{fig3} are approximatively given by the Jaynes-Cummings and Tavis-Cummings 
Rabi frequencies (see subsection II B) as follows: $T_0=2\pi/\Omega_0$, $T_1=4\pi/{\tilde\Omega}_1=\sqrt{2}T_0$ and 
$T_2=4\pi/\tilde{\Omega}_2=\sqrt{2/3}T_0$. 
 This effect can be viewed as a back-action of the `measured' subsystem $S_b$ on the `driving' subsystem $S_a$.

\begin{figure}[t]
\includegraphics[angle=0,width=0.475\textwidth]{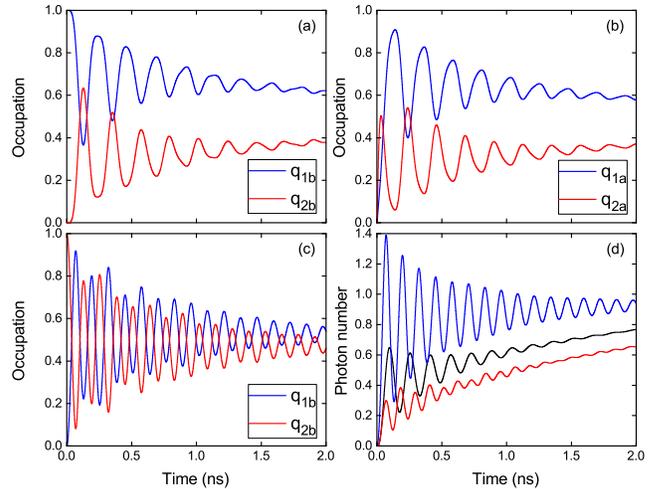}
 \caption{(Color online) The transient Rabi oscillations of the charge occupations $q_{is} $ for different initial states of the
closed system $S_b$. a) $q_{ib}$ for the initial state $|\uparrow_{1b}\rangle$. b) $q_{ia}$ for the initial state $|\uparrow_{1b}\rangle$.
The out-of-phase oscillations are due to correlated processes of emission (absorption) of one photon by $S_a$ and
absorption (emission) by the electron on the lowest level in $S_b$. c) $q_{ib}$ for the initial state $|\uparrow_{2b}\rangle$.
d) The mean photon number ${\cal N}$ for the three initial states: black - $|0\rangle$, red - $|\uparrow_{1b}\rangle$,
blue - $|\uparrow_{2b}\rangle$. Other parameters: $\mu_{La}=45$ meV, $\mu_{Ra}=35$ meV, $g_0=13.25$ $\mu$eV, $\kappa=0$.}
 \label{fig4}
 \end{figure}

To explain this behavior we recall that the charge occupations $q_{ia}$ 
(which  contribute to the current through $S_a$) are optically correlated to the occupations of the closed system $q_{ib}$. 
Then we compared the Rabi oscillations of $q_{ia}$ and $q_{ib}$ to the time-dependent currents. 
When the closed system $S_b$ is not optically active (i.e. if it is either empty or set in the off-resonant regime w.r.t.\ the 
cavity mode) we have already seen that the occupations of the upper system $S_a$ display transient Rabi oscillations with 
period $T_0$ (see Fig.\ \ref{fig2} in subsection III A). 

If $S_b$ is initialized in the state $|\uparrow_{1b}\rangle$ the photons emitted by $S_a$, where the electrons enter via 
sequential tunneling, activate the correlations between the two subsystems. These are confirmed by  Figs.\ \ref{fig4}(a) 
and (b) which present out-of-phase oscillations of the occupations $q_{ia}$ and $q_{ib}$. The oscillations in the charge 
occupations coincide to the ones in the output current from $S_a$. Finally, if subsystem  $S_b$ is prepared in the excited state 
$|\uparrow_{2b}\rangle$ the charge occupations follow the dynamics of the $N=2$ TC subspace so the oscillation period of 
$q_{ib}$ (see Fig.\ \ref{fig4}(c)) roughly decreases by a factor of $\sqrt 3$ when compared to the $N=1$ case. As expected, 
the charge oscillations of $S_a$ are quickly dumped by the tunneling processes, while the oscillations in $q_{ib}$ are clearly 
visible even at longer times.  

The mean photon number is also shown in Fig.\ \ref{fig4}(d) for the three initial states of the closed system. The oscillation periods 
corresponding to the $N=1$ TC subspace is $T_0/\sqrt{2}$, half the period of the charge occupations. 
On the other hand for $N=2$ the mean photon number oscillates with the same period as $q_{ib}$. The different periods for $q_{is}$ and 
${\cal N}$ is essentially due to the fact that the $N=1$ manifold contains only three dressed-states (see Eq.\ (\ref{dressed-TC23})). 
In this case the unitary dynamics of the closed system in the `free' basis shows that the population of the ground-state 
configurations $G_{\sigma\sigma',1}$ (which gives the only contribution to the mean photon number) follows the dynamics 
associated to the larger gap ${\cal E}_{\sigma\sigma'}^{(2)}-{\cal E}_{\sigma\sigma'}^{(3)}$. In contrast, the charge occupations 
obey a slower dynamics given by the smaller gap ${\cal E}_{\sigma\sigma'}^{(3)}-{\cal E}_{\sigma\sigma'}^{(1)}$ such that the 
oscillation period is twice the one of the photon number.  

Since each oscillation period of $J_{Ra}$ is associated to one initial state of $S_b$ the analysis of the transient 
current provides an all-electrical probing of the closed subsystem. Let us note that the steady-state values of the currents shown in 
Figs.\ \ref{fig3}(a)-(c) do not offer any hint on the initial state of $S_b$ or on the $N$-photon subspaces involved in transport. 

\begin{figure}[t]
\includegraphics[angle=0,width=0.475\textwidth]{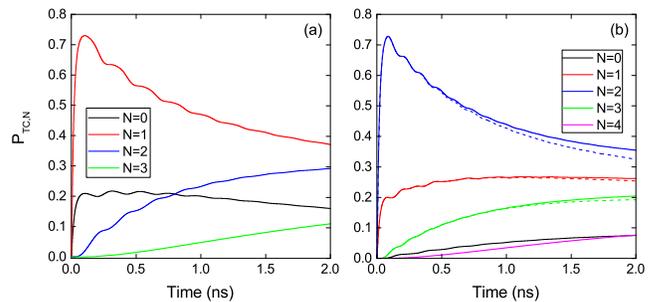}
 \caption{(Color online) The populations of the Tavis-Cummings dressed states with $N=0,1,2,3$ excitations for two initial 
states of the closed system $S_b$. (a) $|\uparrow_{1b}\rangle$, (b) $|\uparrow_{2b}\rangle$. In Fig.\ \ref{fig5}(b) we 
add (see the dashed lines for $N=1,2,3$ and the solid line for $N=4$) the populations obtained by suitably increasing the 
number of photonic Fock states taken into account in the numerical simulations. The parameters are the same as in 
Fig.\ \ref{fig4}.}
 \label{fig5}
 \end{figure}

A valuable insight on the system dynamics is provided by the time-dependent populations $P_{TC,N}$ associated to 
$N$-excitation Tavis-Cummings dressed-states (see Eq.\ (\ref{P_TC_total})). If we switch the coupling to the leads 
with $S_b$ in its single-particle ground state the transient regime (see Figs.\ \ref{fig5}(a)) is mostly described by 
single-excitation dressed states whose total population peaks up to 0.73 at short times and then slowly decreases 
towards the steady-state value. Higher excited states with $N=2$ slowly emerge in the system dynamics as the generation 
of more than one photon becomes possible. One also notices that the total population of the spin-degenerate ground states 
$|G_{\sigma\sigma'},0\rangle$ saturates quite fast at 0.2 and the contribution of the $N=3$ configurations is negligible. One can
therefore argue that the Rabi oscillation period of the transient current must correspond to the TC frequency $\tilde\Omega_1$.
 
If $S_b$ is initially in the state $|\uparrow_{2b}\rangle$ the two-excitation configurations dominate the transient regime 
(see Fig.\ \ref{fig5}(b)), which again support the finding of the Rabi oscillation period given by $\tilde\Omega_2$. 
The single-excitation states, although not negligible, leave no clear fingerprint on the charge dynamics. 
The total population associated to the dressed states in Jaynes-Cummings sector does not exceed 0.06 and therefore were not shown.

From Figs.\ \ref{fig5}(a) and (b) we also notice that the double-emitter system is described by a mixture of 
Tavis-Cummings dressed-states living in subspaces with different excitation numbers $N$. This feature is never encountered in 
the absence of the leads, because the electron-photon interaction is block-diagonal w.r.t.\ the excitation number $N$.  
We find out that the coexistence of states with different excitation number is due to the interplay of the photon emission 
and tunneling in and out processes. Let us consider two possible 'paths' leading to a current in the upper system $S_a$. 
For simplicity we discuss elementary tunneling and optical processes involving the free states (we selected 
for simplicity a single spin orientation $\sigma$ but more complicated combinations are also possible):   
\begin{eqnarray}\nonumber
A:|\sigma_{2a}\sigma_{1b},0\rangle\rightarrow|\sigma_{1a}\sigma_{1b},1\rangle\rightarrow|\sigma_{1b},1\rangle\rightarrow|\sigma_{2a}\sigma_{1b},1\rangle,
\\\nonumber
B:|\sigma_{2a}\sigma_{2b},0\rangle\rightarrow|\sigma_{2a}\sigma_{1b},1\rangle\rightarrow|\sigma_{1b},1\rangle\rightarrow|\sigma_{1a}\sigma_{1b},1\rangle.
\end{eqnarray}
On path A an electron tunnels first on the upper level of subsystem $S_a$ while $S_b$ is on its ground state such that the hybrid system is 
in the state $|\sigma_{2a}\sigma_{1b},0\rangle$. Then a photon is emitted by $S_a$ via electron relaxation on $\varepsilon_{1a}$. Next, the 
same electron tunnels to the right lead and the open system is charged by another electron which now occupies again the upper level 
$\varepsilon_{2a}$. By inspecting Eqs.\ (\ref{G_ssp}-\ref{S_ssp}) we notice that the states $|\sigma_{2a}\sigma_{1b},0\rangle$ and 
$|\sigma_{1a}\sigma_{1b},1\rangle$ are the building block of the $N=1$ dressed states $|\varphi^{(\alpha)}_{1,\sigma\sigma}\rangle$  while 
the `final' state $|\sigma_{2a}\sigma_{1b},1\rangle$ contributes to $N=2$ states $|\varphi^{(\alpha)}_{2,\sigma\sigma}\rangle$. Clearly, 
the path A is more likely when $S_b$ is set to the ground state $|\uparrow_{1b}\rangle$. 

In turn, the 2nd path B is more likely when the closed subsystem $S_b$ is in the excited state $|\uparrow_{2b}\rangle$. In this case a 
tunneling process on the excited level of $S_a$ is followed by a photon emission in the closed subsystem $S_b$. Then as before a double tunneling
involving both leads brings the system to $|\sigma_{1a}\sigma_{1b},1\rangle$. 

In Fig.\ \ref{fig5}(b) we also show the populations of the Tavis-Cummings configurations calculated with a larger number of 
Fock states (i.e.\ the number of Fock states taken into account when solving the GME increases to $N_{{\rm ph}}=5$). One notices that new 
TC configurations with $N=4$ excitations are slowly and rather poorly populated while the configurations with $N=1,2,3$ 
decrease slightly (see the dashed lines in Fig.\ \ref{fig5}(b)).
The populations $P_{TC,N=5}$ and $P_{TC,N=6}$ turn out to be very small and were not shown. Note also that in the time
range where one can read the initial state of the closed subsystem $S_b$ from the Rabi oscillations of the transient current
(i.e.\ for $t<1$ ns) the addition of more photonic states does not lead to a noticeable change of $P_{TC,N}$ for $N=0,3$.

The correspondence between the transient ROs and the initial state of the closed system was recovered for other values of 
the electron-photon coupling strength $g_0$; this is expected as the ratio between the period of the ROs does not depend on $g_0$.
We also find that at fixed value of $g_0$ the number of clear ROs decreases at stronger coupling to the leads when the tunneling time 
becomes smaller than the Rabi periods. As for the effects of cavity losses, the above results are expected to hold as long as $\kappa$ is 
much smaller than the electron-photon coupling strength In particular, the period of the transient Rabi oscillations does 
not depend on $\kappa$.  

Let us comment a bit more on the  Coulomb interaction effects. If the chemical potentials  $\mu_{Ls}$ are pushed upwards more 
complicated transitions between many-body configurations come into play. For example, the excited triplet or singlet states 
can relax to the two-particle ground state by photon emission at suitable frequency of the cavity. Then one expects 
corresponding Rabi oscillations in the transient currents and similar results. On the other hand, by adjusting the 
frequency to these new transitions one detunes the previous transitions between the single-particle levels. 
We also stress that in our calculations the initial state of the hybrid system corresponds to the vaccum field so the mechanism
behind the revival of the population inversion is not activated. Moreover, here we are dealing with
an open system and even if the initial photon configuration would be a superposition of different Fock states,
the dissipation due to the sequential electron tunneling prevents the revival phenomenon.

In a recent experiment \cite{PhysRevApplied.9.014030} the photons emitted by a voltage-biased double QD embedded in a microcavity 
are used to `read' the charge state of a second unbiased double QD placed at the other side of the cavity. 
The `reading' operation is performed by measuring the optical output power or the emission of the biased dot. 
In this context, our theoretical calculations predict the possible {\it electrical} readout of a `target' subsystem by the transient 
current of the `probe' subsystem. 

\section{Conclusions}

The transient and steady-state transport properties of parallel quantum conductors embedded in a photon cavity 
have been calculated from the generalized master equation. The system is described by a Tavis-Cummings model 
suitably extended for interacting and open systems. We propose two settings which exhibit clear effects of the 
coupling between the two conductors via cavity photons. A steady-state effect consists in the removal 
of a Coulomb blockade from one subsystem when a current passes through the second subsystem. This coincides with the switching 
between the Jaynes-Cummings dynamics of a single subsystem to the cavity-mediated Tavis-Cummings dynamics of the double  system.

In a second setup we show that the photon-induced correlations provides information on the initial state of a closed subsystem by 
looking at the transient current which passes through a neighbor open subsystem. More precisely, the period of the Rabi 
oscillations of the `probing' current depends of the initial state of the `probed' system. Note that the `reading' of 
the closed system via photon-mediated interactions is far from being similar to the charge sensing based on mutual 
Coulomb interaction. The back-action effect reported here requires that the levels of the closed system are 
optically active, their gap roughly matching the frequency of the cavity mode. 

The interplay of photon emission/absorption processes and in/out sequential tunneling allows the simultaneous population of 
dressed states with different excitation number. 


\begin{acknowledgments}
The authors acknowledge financial support from CNCS - UEFISCDI grant PN-III-P4-ID-PCE-2016-0221, 
from the Romanian Core Program PN19-03 (contract no.\ 21 N/08.02.2019) and from Reykjavik University, 
grant no. 815051. 
This work was partially supported by the Research Fund of the University of Iceland, and the Icelandic Research Fund, 
grant no.\ 163082-051.
\end{acknowledgments}

\bibliographystyle{natbst}
%

\end{document}